\documentclass[reprint,twocolumn,prl,amsmath,amssymb,superscriptaddress,notitlepage,nofootinbib,longbibliography]{revtex4-1}

\usepackage{xcolor}
\usepackage[export]{adjustbox}
\usepackage{lipsum}
\usepackage{hyperref}
\usepackage{graphicx}  
\usepackage{amsmath, amsfonts, amssymb}
\usepackage{braket}
\usepackage{dsfont}
\usepackage{upgreek}
\usepackage{qcircuit}
\usepackage{tikz}
\usetikzlibrary{matrix}

\usepackage[utf8]{inputenc}
\usepackage[english]{babel}
\usepackage{hyperref}
\hypersetup{
    colorlinks=true,
    linkcolor=black,
    filecolor=black,  
    citecolor=black,
    urlcolor=black,
}
\usepackage{array}
\usepackage{mathtools}
\usepackage{booktabs}
\usepackage{siunitx}

\newcommand{\adag}{\hat{a}^{\dagger}}
\newcommand{\bdag}{\hat{b}^{\dagger}}

\DeclarePairedDelimiter{\abs}{\lvert}{\rvert}

\newcolumntype{C}{>{$}c<{$}}
\AtBeginDocument{
    \heavyrulewidth=.08em
    \lightrulewidth=.05em
    \cmidrulewidth=.03em
    \belowrulesep=.65ex
    \belowbottomsep=0pt
    \aboverulesep=.4ex
    \abovetopsep=0pt
    \cmidrulesep=\doublerulesep
    \cmidrulekern=.5em
    \defaultaddspace=.5em
}

\begin{document}

\title{Programmable photonic system for quantum simulation in arbitrary topologies}

\author{Ben Bartlett}
\affiliation{Department of Applied Physics, Stanford University, Stanford, California 94305, USA}
\affiliation{Now at PsiQuantum, Palo Alto, California 94304, USA}

\author{Olivia Y. Long}
\affiliation{Department of Applied Physics, Stanford University, Stanford, California 94305, USA}

\author{Avik Dutt}
\affiliation{Department of Mechanical Engineering, University of Maryland, College Park, Maryland 20742, USA}
\affiliation{Institute for Physical Science and Technology, University of Maryland, College Park, Maryland 20742, USA}

\author{Shanhui Fan}
\email{shanhui@stanford.edu}
\affiliation{Department of Electrical Engineering, Stanford University, Stanford, California 94305, USA}

\begin{abstract}
Synthetic dimensions have generated great interest for studying many types of topological, quantum, and many-body physics, and they offer a flexible platform for simulation of interesting physical systems, especially in high dimensions. In this Letter, we describe a programmable photonic device capable of emulating the dynamics of a broad class of Hamiltonians in lattices with arbitrary topologies and dimensions. We derive a correspondence between the physics of the device and the Hamiltonians of interest, and we simulate the physics of the device to observe a wide variety of physical phenomena, including chiral states in a Hall ladder, effective gauge potentials, and oscillations in high-dimensional lattices. Our proposed device opens new possibilities for studying topological and many-body physics in near-term experimental platforms.
\end{abstract}
\maketitle

The emerging concept of synthetic dimensions in photonics has generated great interest for topological physics \cite{Dutt2020ADimensions, yuan_synthetic_2018, baum_setting_2018, price_roadmap_2022, ozawa_topological_2019}, optimization \cite{inagaki_coherent_2016, mcmahon_fully_2016, marandi_network_2014, novel_neural_network_temporal_synth_dim_peng_arxiv_2021}, and quantum simulation and computation \cite{boada_quantum_2012, ozawa_synthetic_2016, Chalabi2019SyntheticWalks, Bartlett2021DeterministicDimension, pysher_parallel_2011, roslund_wavelength-multiplexed_2014, yoshikawa_invited_2016, larsen_deterministic_2019}. Synthetic dimensions are formed by controlling couplings between degrees of freedom of a system, either by repurposing the usual geometric dimensions, such as space \cite{lustig_photonic_2019} or time \cite{regensburger_paritytime_2012, wimmer_experimental_2017, marandi_network_2014, inagaki_coherent_2016, mcmahon_fully_2016, leefmans_topological_2022, tiurev_fidelity_2021, hilaire_near-deterministic_2022}, or by augmenting these dimensions with internal degrees of freedom, such as frequency~\cite{yuan_photonic_2016, ozawa_synthetic_2016, bell_spectral_2017, hu_realization_2020, wang_multidimensional_2020}, spin~\cite{celi_synthetic_2014, mancini_observation_2015, stuhl_visualizing_2015, Dutt2020ADimensions}, or orbital angular momentum~\cite{luo_quantum_2015, yuan_photonic_2019}. Since couplings in synthetic dimensions can be dynamically reconfigured and are not fixed by a physical structure, one can scalably implement high-dimensional lattices with complex topologies, making this an ideal platform for quantum simulation.

In this Letter, we describe a programmable photonic device capable of simulating the dynamics of interacting bosons in lattices with arbitrary dimensions, topologies, and connectivities using a synthetic time dimension. A large class of prototypical condensed matter Hamiltonians can be described by local two-body interactions on an arbitrary lattice. This class of Hamiltonians, which includes tight-binding models, the Hubbard and Bose-Hubbard models and their various extensions~\cite{dutta_non-standard_2015}, and the Harper-Hofstadter-Hubbard model~\cite{Repellin2020FractionalProfiles}, can in general be described as (using $\hbar=1$ throughout this Letter):
\begin{multline}
\label{eq:hubbard_hamiltonian}
\hat{H} = -
\sum_{\langle m, n \rangle} \left( \kappa_{mn} e^{i \alpha_{mn}} \hat{a}^\dagger_m \hat{a}_n + \mathrm{H.c.} \right) \\
+ \mu \sum_m \hat{a}^\dagger_m \hat{a}_m 
+ U \sum_{m} \hat{a}^\dagger_m \hat{a}^\dagger_m \hat{a}_m \hat{a}_m,
\end{multline}
where $\kappa_{mn}$ and $\alpha_{mn}$ respectively denote the tunneling coefficients and phases between connected sites $\langle m,n \rangle$, $\hat{a}^\dagger_{m}$ creates a boson at site $m$, $\mu$ is the chemical potential, and $U$ is the Hubbard interaction strength. The first term describes the tunneling of a particle between sites $m$ and $n$, with a complex tunneling strength with amplitude $\kappa_{mn}$ and phase $\alpha_{mn}$; the second term sets the energy per particle $\mu$; the third term is an on-site interaction potential with strength $U$ which is active when a site contains more than one particle. This very general class of Hamiltonians exhibits rich phase diagrams and relates to quantum magnetism, high-temperature superconductors, and magnetic insulators, among many other applications. \cite{auerbach_interacting_electrons, Zapf2014Bose-EinsteinMagnets, Giamarchi2008Bose-EinsteinInsulators}

\begin{figure*}[t]
\centering 
\includegraphics[width=\textwidth]{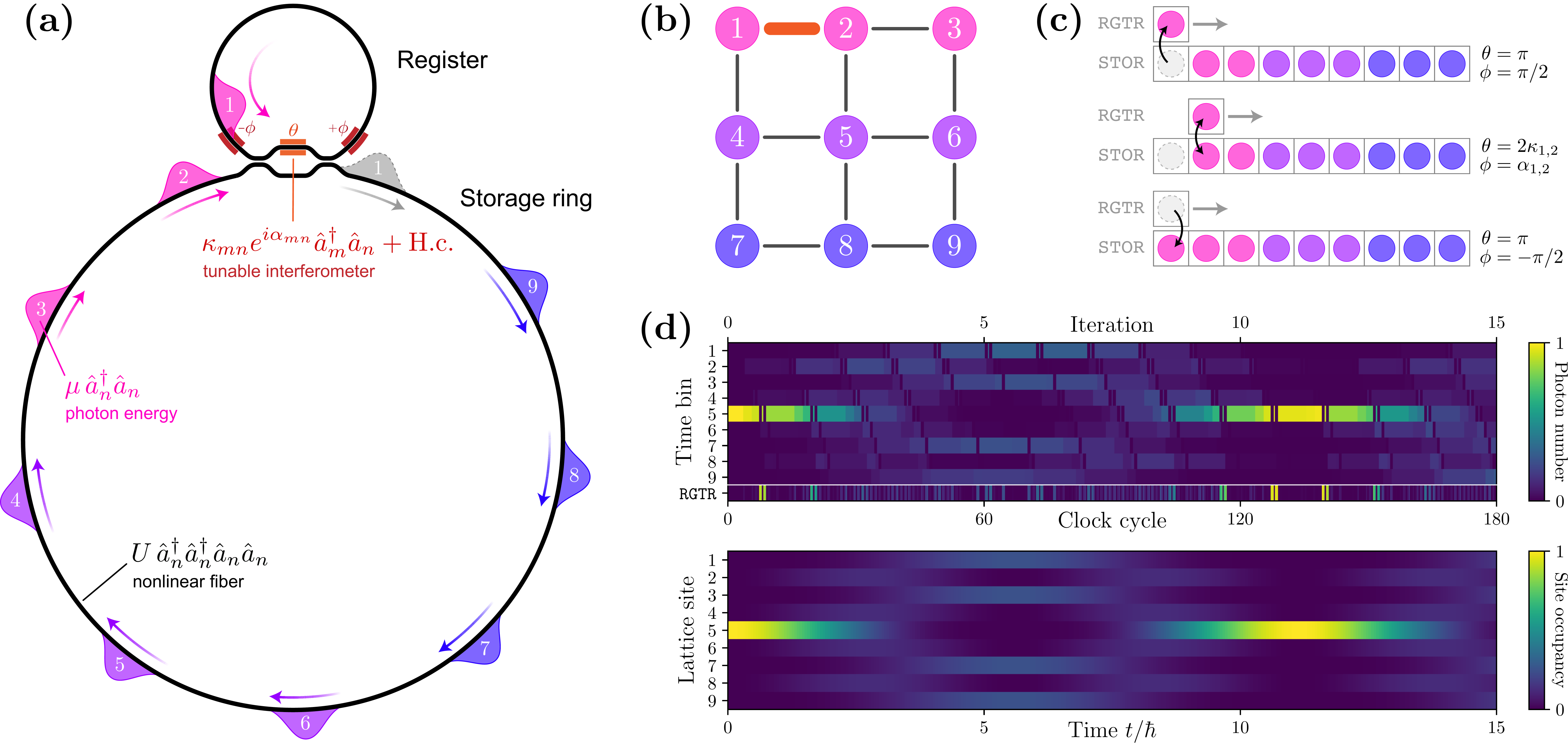}
\caption{
Architecture for the programmable photonic quantum emulator described in this Letter.
\textbf{(a)} 
The physical design of the device. Classical laser pulses or single-photon pulses propagate clockwise through a fiber storage ring. A programmable Mach-Zehnder interferometer connects the storage ring to a register loop which has an optical path length $\Delta x$ equal to the length of a single time bin. By setting the phase shift values in the MZI, the hopping coefficients and phases $\kappa_{mn}$, $\alpha_{mn}$ can be programmatically adjusted. Photons have energy $\mu \equiv \hbar \omega_0$, and by using a $\chi^{(3)}$-nonlinear fiber, a nonlinear interaction potential $U$ can be emulated. 
\textbf{(b)} 
An example 2D grid lattice to be emulated by the device. Node labels correspond to photon pulse indices, and the device as shown in panel (a) is in the process of constructing the orange edge connecting nodes 1 and 2 with $(\kappa_{1,2}, \alpha_{1,2})$.
\textbf{(c)} 
Illustration of a single clock cycle of the emulator constructing the interaction $(\kappa_{1,2}, \alpha_{1,2})$ in three steps. First, phase shifters are set to transfer photon 1 into the register. Second, photon 1 is interacted with photon 2 using $\theta = 2 \kappa_{1,2}$ and $\phi = \alpha_{1,2}$. Third, the pulse (which now may contain a mixture of photons 1 and 2) is returned to its original time bin.
\textbf{(d)}
The evolution of the state of the device while emulating a tight-binding Hamiltonian over the lattice shown in panel (b). The bottom panel depicts the exact evolution of the target Hamiltonian over time, while the top panel shows the state of the emulator at each clock cycle, including register swaps and intermediate states between full iterations. A large value of $\kappa = 0.2$ was used for visual clarity, but more accurate results may be obtained by using smaller $\kappa$ and running the emulation for a commensurately longer wall-clock time.
}
\label{fig:design}
\end{figure*}

We propose a system that emulates the dynamics of the Hamiltonian in Eq. \ref{eq:hubbard_hamiltonian} using a synthetic temporal dimension. The design consists of a waveguide loop exhibiting a Kerr nonlinearity, which we refer to as the ``storage ring'', in which a train of single-photon pulses propagates in a single direction, with each pulse occupying its own time bin. A second loop, the ``register'', is connected to the storage ring using a Mach-Zehnder interferometer (MZI) with two tunable phase shifters, $\theta$ and $\phi$. The hardware of the device is chosen to emulate each term of the Hamiltonian with dedicated components. The first term of Eq. \ref{eq:hubbard_hamiltonian} is implemented by the tunable MZI; the second term arises naturally from the total photon energy in each time bin; the two-photon potential in the third term results from using a Kerr-nonlinear fiber for the storage and register loops. We will briefly derive how each component implements the desired behavior and then describe how to program the device.

A system evolving for a time interval $t$ under the Hamiltonian given in Eq. \ref{eq:hubbard_hamiltonian} has a propagator $e^{-i \hat{H} t}$. We can split the exponential of the summation into a product of exponentials to within $\mathcal{O}(\kappa^2 + \kappa U \cos \alpha)$, where $\kappa$ and $\alpha$ are typical values of $\kappa_{mn}, \alpha_{mn}$ (see Supplementary Materials for a more detailed derivation \cite{SM}):
\begin{widetext}
\begin{align}
\begin{split}
\label{eq:propagator_split}
e^{-i\hat{H}t} 
&= \exp \left[ -it \left(-\sum_{\langle m, n \rangle} \kappa_{mn} \left( e^{i \alpha_{mn}}  \hat{a}^\dagger_m \hat{a}_n  + e^{-i \alpha_{mn}} \hat{a}^\dagger_n \hat{a}_m \right) + \mu \sum_m \hat{a}^\dagger_m \hat{a}_m + U \sum_{m} \hat{a}^\dagger_m \hat{a}^\dagger_m \hat{a}_m \hat{a}_m \right)\right] \\
&\approx \left[\prod_{\langle m, n \rangle} \exp\left(i \kappa_{mn} \left(e^{i \alpha_{mn}}  \hat{a}^\dagger_m \hat{a}_n + e^{-i \alpha_{mn}} \hat{a}^\dagger_n \hat{a}_m \right)\right) 
\right]^t \times e^{-it \left( \mu \sum_m \hat{a}^\dagger_m \hat{a}_m + U \sum_m \hat{a}^\dagger_m \hat{a}^\dagger_m \hat{a}_m \hat{a}_m \right)}.
\end{split}
\end{align}
\end{widetext}

We therefore have a propagator that is a product of two parts: a continuous time evolution term $e^{-it \left( \mu \sum_m \hat{a}^\dagger_m \hat{a}_m + U \sum_m \hat{a}^\dagger_m \hat{a}^\dagger_m \hat{a}_m \hat{a}_m \right)}$, which arises naturally from the photon energy per time bin ($\mu$) and Kerr nonlinearity of the fiber ($U$), and the $\exp(i\kappa_{mn} e^{i\alpha_{mn}} \hat{a}^\dagger_m \hat{a}_n + \mathrm{H.c.})$ terms, which are implemented in discrete time evolution by a sequence of passes through the tunable MZI. We now show how the device physics emulates the dynamics of the propagator.

For the chemical potential term, we can write the Hamiltonian for a photon with an arbitrary spectrum as $\hat{H}_\mathrm{EM} =\int dk \sum_m \frac{1}{2} \hbar \omega_k \left(\hat{a}_{m,k}^\dagger \hat{a}_{m,k} + \hat{a}_{m,k} \hat{a}_{m,k}^\dagger \right)$. If we can assume that the photons are spectrally narrow about a carrier frequency $\omega_0$, we can approximate this as $\hat{H}_\mathrm{EM} \approx \frac{1}{2} \hbar \omega_0 \sum_m \left(\hat{a}_{m}^\dagger \hat{a}_{m} + \hat{a}_{m} \hat{a}_{m}^\dagger \right) = \hbar \omega_0 \sum_m \left(\hat{a}_{m}^\dagger \hat{a}_{m} + \frac{1}{2}\right) \equiv \mu \sum_m \hat{a}_{m}^\dagger \hat{a}_{m}$, which directly gives us the desired chemical potential term.

The nonlinear potential naturally arises from the use of a nonlinear fiber. Consider a section of a Kerr-nonlinear fiber corresponding to one time bin, with length $\Delta x$ and volume $V$. The material polarization at frequency $\omega$ induced by an electric field $E(\omega)$ is given by $\mathcal{P}_\mathrm{NL} (\omega) = 3 \varepsilon_0 \chi^{(3)} (\omega) \abs{E(\omega)}^2 E(\omega)$, where $\chi^{(3)}$ is the third-order susceptibility tensor, which can be treated as a scalar for isotropic media such as glass. The energy density $\mathcal{U}_\mathrm{NL}$ is related as $\mathcal{P}_\mathrm{NL}  = \partial \mathcal{U}_\mathrm{NL} / \partial E^*$, and the Hamiltonian of this system, again assuming a narrow bandwidth about $\omega_0$, is $\hat{H}_\mathrm{NL} = \int_V \mathcal{U} (\omega_0) d^3 \vec{r}$. After quantizing the field amplitudes as $E(\omega_0) \mapsto \sqrt{\frac{\hbar \omega_0}{2 \epsilon_0 V}} \left(\hat{a}^{\dagger}_{k_0} e^{+i (\omega_0 t - k_0 z) } + \mathrm{H.c.} \right)$ and transforming into real space, we obtain $\hat{H}_\mathrm{NL} = \left( \frac{9 \hbar^2 \omega_0^2}{8 \epsilon_0 n_0^4 V^2} \int_V \chi^{(3)}  d^3\mathbf{r} \right) \hat{a}^\dagger \hat{a}^\dagger \hat{a} \hat{a} + C \equiv U \hat{a}^\dagger \hat{a}^\dagger \hat{a} \hat{a} + C$, where the nonlinear potential coefficient is $U=\frac{9 \hbar^2 \omega_0^2}{8 \epsilon_0 n_0^4 V^2} \int_V \chi^{(3)}  d^3\mathbf{r}$ and where $C$ is some constant corresponding to an overall energy shift. Applying this to each time bin gives us the desired $U \sum_m \hat{a}^\dagger_m \hat{a}^\dagger_m \hat{a}_m \hat{a}_m$ nonlinear potential term.

Finally, the hopping terms arise from programmatically modulating the phase shifts in the MZI. To interfere two photons $m$ and $n$ with strength $\kappa_{mn}$ and phase shift $\alpha_{mn}$, the basic idea is to swap pulse $m$ into the register ring, wait for pulse $n$ to reach the MZI, interfere the pulses, then return pulse $m$ to the storage ring when time bin $m$ cycles back. Consider the MZI shown in Figure \ref{fig:design}(a) with phase shifters $\pm \phi$ and $\theta$. Define bosonic mode operators $\hat{a}^\dagger _n, \hat{a}^\dagger_0$ and $\hat{b}^\dagger _n, \hat{b}^\dagger_0$, which create a photon in time bin $n$ or time bin $0$ (the register bin), and at the input or output of the MZI, respectively. We can relate the output and input mode operators as:
\begin{align}
\begin{split}
\label{eq:transfer_matrix}
\begin{bmatrix} 
\bdag_0 \\ 
\bdag_n
\end{bmatrix} 
&= 
\begin{pmatrix}
\cos \frac{\theta}{2} & i e^{i \phi} \sin \frac{\theta}{2} \\
i e^{-i\phi} \sin \frac{\theta}{2} & \cos \frac{\theta}{2}
\end{pmatrix}
\begin{bmatrix} 
\adag_0 \\ 
\adag_n
\end{bmatrix} 
\\
&= \exp\left[ i \frac{\theta}{2} \left(e^{i\phi} \adag_0 \hat{a}_n + e^{-i\phi} \adag_n \hat{a}_0\right) \right]
\begin{bmatrix} 
\adag_0 \\ 
\adag_n
\end{bmatrix}
\\
&\equiv \hat{M}_{0,n} \left(\theta, \phi \right) 
\begin{bmatrix} 
\adag_0 \\ 
\adag_n
\end{bmatrix}.
\end{split}
\end{align}
It is easily verified that the following identity holds: $\hat{M}_{0,m} \left(\pi, -\pi/2 \right)   \hat{M}_{0,n} \left(\theta, \phi \right)   \hat{M}_{0,m} \left(\pi, +\pi/2 \right) = \exp\left[ i \theta/2 \left( e^{i \phi}  \hat{a}^\dagger_m \hat{a}_n  + e^{-i \phi} \hat{a}^\dagger_n \hat{a}_m \right)\right] \equiv \hat{T}_{m,n} \left(\theta, \phi\right)$. If we define $\kappa \equiv \theta/2$ and $\alpha \equiv +\phi$, we obtain the transfer matrix:
\begin{equation}
\hat{T}_{mn}\left(\kappa, \alpha\right) = \exp\left[ i \kappa \left( e^{i \alpha}  \hat{a}^\dagger_m \hat{a}_n  + e^{-i \alpha} \hat{a}^\dagger_n \hat{a}_m \right)\right].
\end{equation}
The middle $\theta$ phase shifter thus allows us to control the strength of the coupling $\kappa$, while the outer phase shifters $\pm \phi$ control the hopping phases.

By performing this sequence of passes through the MZI $\hat{T}_{\langle m, n \rangle} \equiv \prod_{\langle m,n \rangle} \hat{T}_{mn} \left(\kappa_{mn}, \alpha_{mn}\right)$ for every photon pair $\langle m, n \rangle$ which corresponds to an adjacent pair of lattice sites $m$ and $n$ in the Hamiltonian, we complete one ``iteration'' of the emulator. If we allow the system to evolve for $t$ iterations, we obtain a total transfer matrix which is exactly the first term in Eq. \ref{eq:propagator_split}:
\begin{equation}
\hat{T}_{\langle m, n \rangle}^{\; t} = \left(\prod_{\langle m, n \rangle} \exp\left[ i \kappa_{mn} \left(e^{i\alpha_{mn}} \adag_m \hat{a}_n + \text{H.c.} \right) \right] \right)^t
\end{equation}

Therefore, all three components of the propagator are present, and the evolution of a state in the device for $t$ iterations is described term-by-term by the propagator in Eq. \ref{eq:propagator_split}. To adjust the relative values of continuous time evolution variables ($\mu, U$) and discrete time evolution variables ($\kappa, \alpha$), one can adjust the photon energies $\mu$, Kerr interaction strength $U$, time bin size $\Delta x$, or phase shifter values $\theta, \phi$.

\begin{figure*}[t]
\centering 
\includegraphics[width=\textwidth]{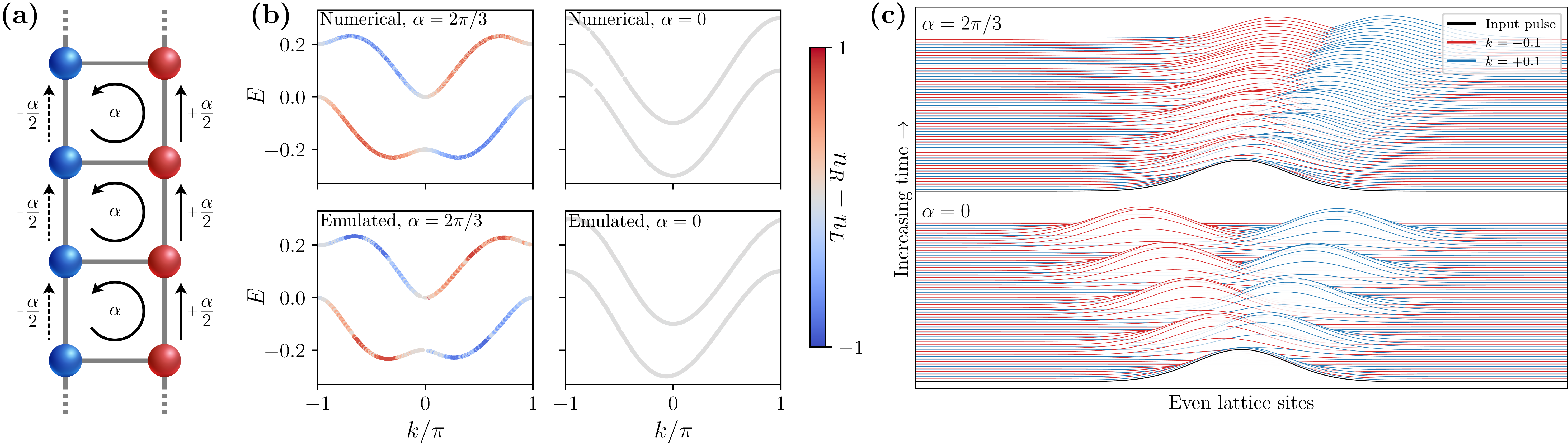}
\caption{
\textbf{(a)} 
Lattice diagram for a two-legged synthetic Hall ladder emulated with the device. By varying the inter-rung hopping phases $\alpha$, an effective controllable magnetic field can be induced in the lattice. 
\textbf{(b)} 
Band structure of the system computed by diagonalizing the Hamiltonian for the exact case (top panels) and as emulated in the device (bottom) in the presence (left) and absence (right) of a synthetic magnetic field. Projection operators to the left and right nodes are color coded for each eigenstate. In all cases the Hamiltonian is represented in real space; for each eigenstate with eigenvalue $E$, we compute $k$ with peak detection of its Fourier transform. This results in small numerical instabilities which are present in both the exact and emulated cases. Other parameters for this simulation: $\kappa = 0.1$, $\alpha = 2\pi/3$ or $\alpha=0$, $\mu = U = 0$, number of lattice sites $D=1000$, number of bosons $N=1$.
\textbf{(c)} 
Experimental signature for the propagation of chiral edge currents on the left leg of the ladder. A Gaussian input state is created with some initial $k=\pm 0.1$ by exciting multiple time bins with a phase difference between bins. When the gauge field is turned off ($\alpha = 0$), the pulses propagate in opposite directions, but when the field is turned on ($\alpha=2\pi/3$), the motion in one direction is inhibited.
}
\label{fig:bandstructure}
\end{figure*}

The programmable MZI can construct lattices with arbitrary topology and connectivities. Consider a Hamiltonian of the form in Eq. \ref{eq:hubbard_hamiltonian} defined over a lattice described by an undirected graph $G = (V, E)$, as shown in Figure \ref{fig:design}(b). We designate a time bin $m$ for each lattice site $m \in V$, and for each edge $e = (m,n) \in E$ which couples sites $m$ and $n$ with coupling strength $\kappa_{mn}$ and hopping phase $\alpha_{mn}$, we perform a sequence of three passes through the MZI to interact time bins $m$ and $n$, as in Figure \ref{fig:design}(c). The first pass $\hat{M}_{0,m} \left(\pi, +\pi/2 \right)$ swaps photon $m$ into the empty register; the second pass $\hat{M}_{0,n} \left(2 \kappa_{mn}, \alpha_{mn} \right)$ performs the interaction between the register and time bin $n$; the third pass $\hat{M}_{0,m} \left(\pi, -\pi/2 \right)$ returns the photon to time bin $m$. This set of operations takes one ``clock cycle'' to complete, which is defined as the time for a pulse to fully propagate once around the storage ring. Constructing all $e \in E$ completes one iteration of the emulator, and the state is allowed to evolve for $t$ iterations. The edges can be constructed in any order as long as $\kappa$ is small, which is always possible to do by decreasing $\kappa, \mu, U$ by some constant factor and running the emulator for a commensurately longer wall-clock time. 
It is also possible to modify the design of this device to include hardware optimizations for specific graphs. For example, the iteration time of a 2D square lattice with $N$ sites can be reduced from $N$ cycles to $2\sqrt{N}$ cycles per iteration by having two register rings of size $N$ and $\sqrt{N}$ to explicitly handle the vertical and horizontal strides of the graph.

To more concretely show the capabilities of our proposal, we now provide several demonstrations of the device emulating systems of interest with experimentally measurable signatures. We show the device can create an effective gauge potential by emulating a synthetic Hall ladder, we demonstrate the quantum nature of the device by trapping a a two-photon state using a synthetic field, and we demonstrate the reconfigurability of the device by emulating the evolution of a Bose-Hubbard Hamiltonian on a four-dimensional tesseract lattice. For these demonstrations, we wrote a Python simulator\footnote{All simulation code for this Letter is available online at \texttt{\hyperlink{https://github.com/fancompute/synthetic-hamiltonians}{github.com/fancompute/synthetic-hamiltonians}}.} built with \texttt{QuTiP} \cite{Johansson2012QuTiP:Systems} which efficiently simulates the detailed physics of the device emulating a system of interest, such as register swaps and time bin interactions, and compares this against the exact Hamiltonian evolution. The simulator represents the state space of the system with a permutationally invariant bosonic lattice representation allowing for tractable simulation of Hamiltonians over moderately large lattices. This simulation method is described in greater detail in the Supplementary Materials \cite{SM}.

Figure \ref{fig:bandstructure} shows an emulated synthetic Hall ladder and obtains a similar band structure as the recent experimental results of Ref. \cite{Dutt2020ADimensions}. This system exhibits chiral edge states in the presence of an effective magnetic field, which is induced by adding translation-invariant hopping phases $\pm \alpha/2$ to the outer edges of the ladder using the MZI. Figure \ref{fig:bandstructure}(a) depicts the emulated ladder system; left and right nodes on each rung are mapped to pulses in even- and odd-indexed time bins. The band structures for the target and emulated Hamiltonians for this system are shown in Figure \ref{fig:bandstructure}(b) for hopping phases $\alpha = 2\pi / 3$ and $\alpha=0$. Chiral edge states are clearly visible in the case of $\alpha = 2\pi/3$, indicating the presence of an effective gauge potential. The propagation of these chiral currents on the left leg of the ladder is shown Figure \ref{fig:bandstructure}(c). In the presence of a gauge field, only one-way motion is allowed. The band structures for the synthetic case are computed by simulating one iteration of the propagator $\hat{G} = e^{-i \hat{H} (t=1)}$ in the device, taking the matrix logarithm $\hat{H} = \frac{\log \hat{G}}{-i}$, and then diagonalizing $\hat{H}$; $k$ values are computed using peak detection of the eigenstate Fourier transform (see Supplementary Materials).

As shown in Fig.~\ref{fig:bandstructure}(b), the band structure from the emulated system (bottom row) closely matches the desired band structure (top row, see also Ref. \cite{hugel_chiral_2014}), as well as the experimental results from very different platforms (Fig. 2 of Ref. \cite{Dutt2020ADimensions}). This shows that the simulation of our device physics faithfully constructs the desired synthetic Hall Hamiltonian. Furthermore, because this demonstration uses only single boson, the single photon pulse could be substituted for a classical laser pulse which could be periodically re-amplified and reshaped, negating much of the experimental concerns related to attenuation and pulse deformation.

\begin{figure}[t]
\centering 
\includegraphics[width=\columnwidth]{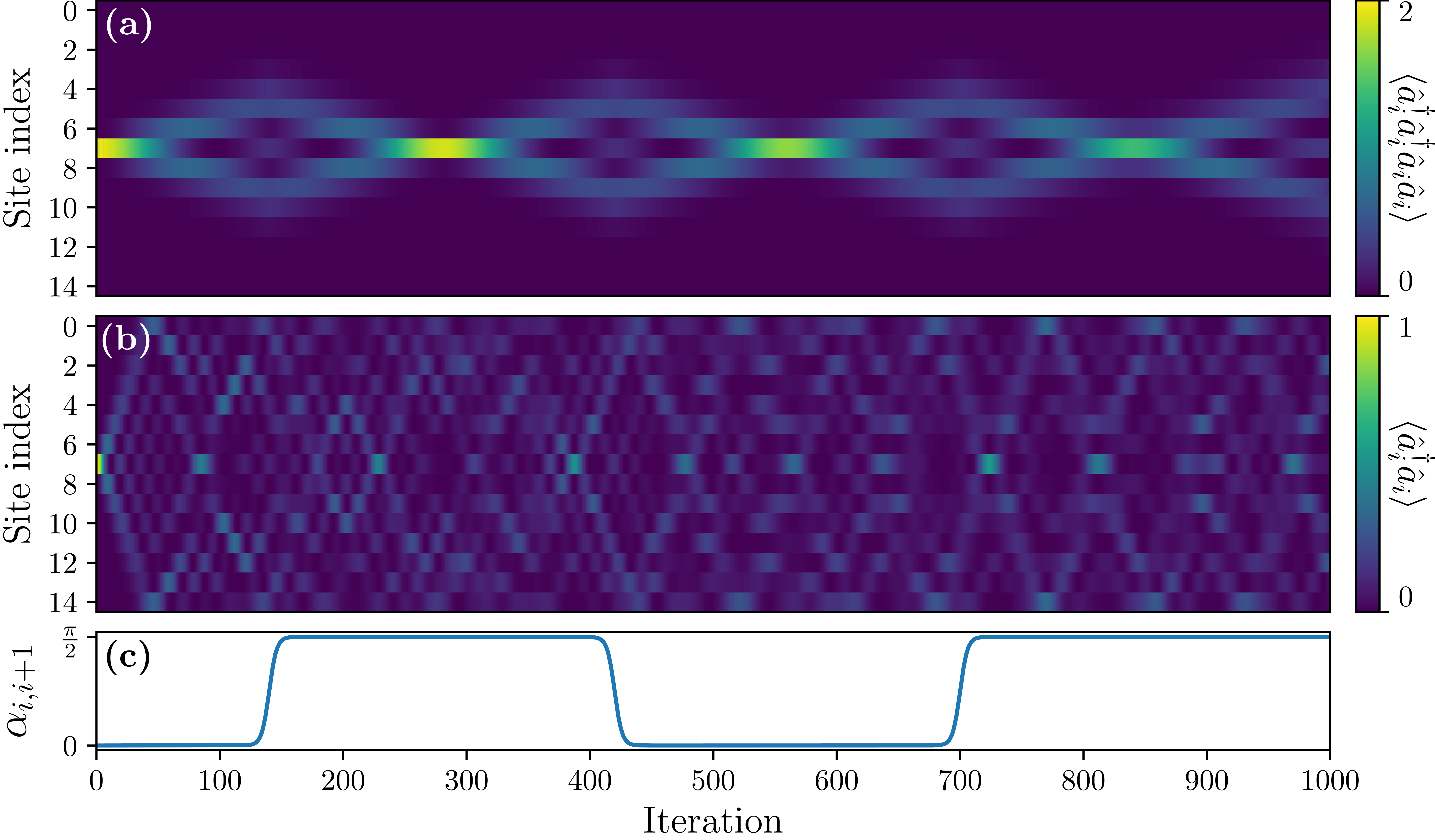}
\caption{
Emulated evolution of \textbf{(a)} a two-photon state and \textbf{(b)} a single-photon state in a 1D lattice as \textbf{(c)} time-dependent hopping phases are varied. The changing hopping phases introduce a changing gauge potential which causes the two-photon state to experience an effective electric field. The single-photon state is unaffected by this field.
}
\label{fig:two_photon_lensing}
\end{figure}

Next, to demonstrate the quantum capabilities of the device, we show how a two-photon state can be manipulated by introducing time-dependent hopping phases $\alpha (t)$ on a 1D lattice while using nonlinearity which is strong compared to the coupling constants $U \gg \kappa$. Figure \ref{fig:two_photon_lensing} depicts the evolution of a two-photon state and a single-photon state under time-dependent hopping phases $\alpha(t)$. The energetic gap between $U \gg \kappa$ means that while $\alpha(t) = 0$, the two-photon state evolves the same as the single-photon state, but with a slower timescale for the evolution.\footnote{The two-photon state in panel (a) undergoes slower evolution because the Hamiltonian has no terms which can transport two photons simultaneously between lattice sites. Thus, evolution is only allowed via single-photon transport through an intermediate state which is lower in energy by $U$. This intermediate state never develops a sizable population because it is off-resonant from the initial and final states.} As $\alpha(t)$ is changed, $\partial \alpha / \partial t$ introduces an effective field, analogous to $\vec{E} = -\nabla V - \partial \vec{A} / \partial t$, which causes the two-photon state to look like it is ``lensing'' back to its original configuration. This field is maximized at odd multiples of $\pi/2$, and by choosing suitable amplitude, duration, and periodicity of $\alpha(t)$, the two-photon state can effectively be trapped in the center of the lattice. The single-photon state is unaffected by the field, since we can perform a gauge transformation of the single-photon basis states as $\hat{a}^\dagger_n \mapsto \hat{b}^\dagger_n e^{i n \alpha(t)}$ which eliminates the effect of $\alpha(t)$.

\begin{figure}[t]
\centering 
\includegraphics[width=\columnwidth]{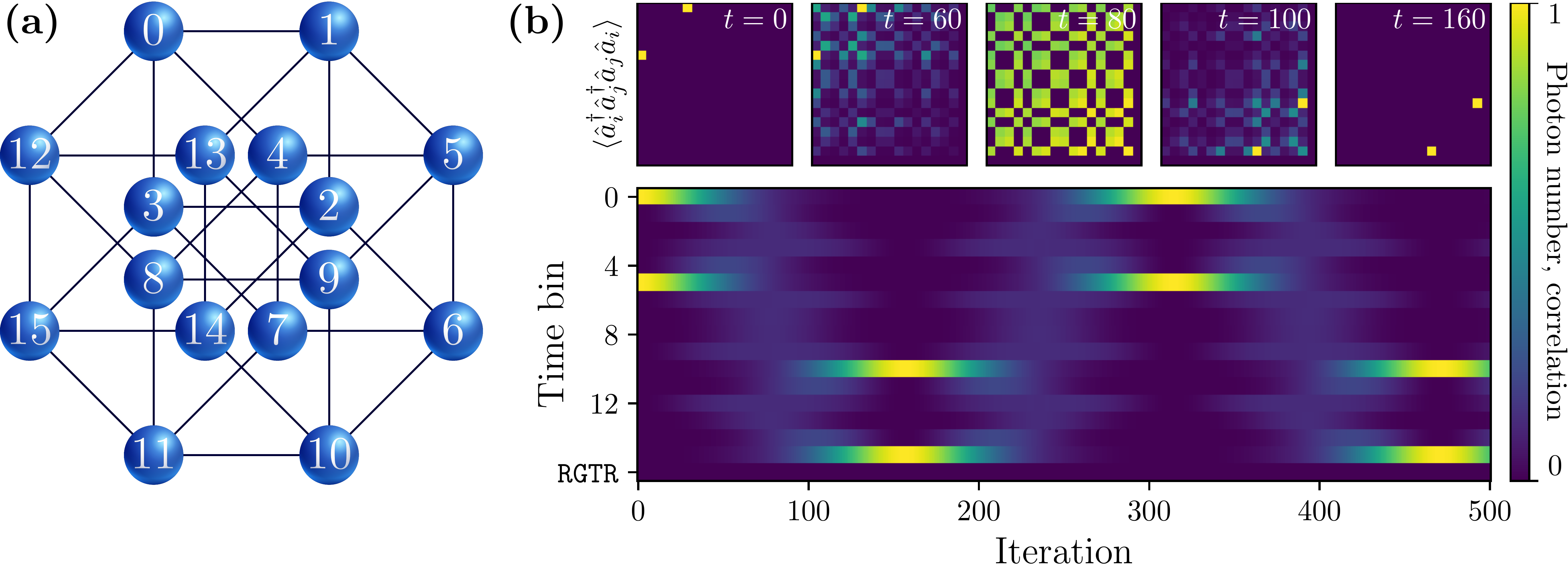}
\caption{
Emulation of a tight-binding Hamiltonian over a four-dimensional tesseract. 
\textbf{(a)} 
Projection of the tesseract graph which defines the lattice. 
\textbf{(b)} 
Evolution of a two-photon state exhibiting oscillations between time bins $0\leftrightarrow 10$ and $5\leftrightarrow 15$. Parameters: $\kappa = 0.01$, $\alpha=\mu=U=0$.
}
\label{fig:tesseract}
\end{figure}

Finally, we demonstrate how the programmable nature of the device allows for emulation of complex, high-dimensional topologies. Figure \ref{fig:tesseract} shows the evolution of a tight-binding Hamiltonian over a four-dimensional tesseract lattice emulated using the device. This demonstration uses a single degree of freedom (time) to emulate four independent physical synthetic dimensions. A projection of the non-planar graph defining the lattice is shown in Figure \ref{fig:tesseract}(a). The evolution of a two-photon state over this tesseract is shown in Figure \ref{fig:tesseract}(b): photons are initially placed in time bins 0 and 5, and oscillations across the tesseract are visible, with the photons oscillating between sites $0\leftrightarrow 10$ and $5 \leftrightarrow 15$. (This is the expected behavior, representing the four-dimensional analogue of a boson oscillating between the corners of a $2 \times 2$ square lattice.) Two-photon correlation matrices are shown at different points in time in the upper panels. The state is plotted at the end of each iteration of the device; since photons have been swapped out of register time bin at the end of each iteration, it is shown to be empty at all times.

Let us now briefly discuss the feasibility of this proposed device in the presence of experimental imperfections. The main limitations of the device are dispersion within the fiber loops\footnote{
Dispersion is unlikely to have an important effect in emulation quality. Typical pulse parameters in a low-dispersion fiber allow for distinguishability over thousands of kilometers of distance (see Supplementary Materials).
}, optical attenuation in the fiber\footnote{
For classically emulable cases (where the total boson number is $N=1$ or where the initial state is well-approximated by a coherent state), single-photon pulses can be replaced by classical laser pulses with complex amplitudes, which can be re-amplified as needed, so attenuation and insertion loss is much less of a concern.
}, phase shifter actuation speed and insertion losses, and limitations on the nonlinear potential $U$. 

For non-classically emulable cases with no nonlinearity ($U=0$), single photon pulses must be used, which cannot be re-amplified, so attenuation and insertion losses will constrain the maximum lattice size which can be emulated with a given fidelity. If we take the tesseract graph with two photons from Figure \ref{fig:tesseract} as an example, using a pulse width of 2 cm and a time bin size of 1 ns, the allowable cycle loss $L$ to emulate a single iteration with $90 \%$ fidelity is $1-L = 0.998$ or $L = -.007 \text{dB / cycle}$. Ignoring MZI insertion losses, this corresponds to $-2.23$ dB/km fiber attenuation, which is easily possible using commercially available fibers (with attenuation as low as $-0.17$ dB/km).

The most difficult cases to emulate are non-classical with large values of $U$. Highly nonlinear photonic crystal fibers filled with a high-density atomic gas can create nonlinearity up to $U / \hbar \sim 1$ GHz in the few-photon regime. \cite{venkataraman_phase_2013, yuan_creating_2020} To compare this to the numerical values of $\kappa, U$ used in the simulations in this work, consider a time bin of size $\Delta t = n_\text{fiber} \Delta x / c$, where $n_\text{fiber}$ is the refractive index. If there are $N$ time bins, then the clock cycle time of the device is $N \Delta t$, so the frequency units for a numerical value of $\kappa = 1.0$ are $\left(2 \pi N \Delta t\right)^{-1}$. For $\kappa = 0.2$ and nine time bins, as used in the lattice for Figure \ref{fig:design}, this corresponds to $\kappa \approx 0.007$ GHz. The value of $U$ is independent of cycle time since it is distributed throughout the length of the fiber ring, and using current nonlinear fibers could be made several orders of magnitude larger than $\kappa$. Furthermore, recent improvements in the demonstrated nonlinearity-to-loss ratios for integrated photonic platforms approaching 1.5\%~\cite{zhao_ingap_2022, lu_toward_2020} (albeit in $\chi^{(2)}$ materials) show promise for achieving even higher values of $U$ in the near future.

In this Letter we have presented a design for a programmable photonic device capable of emulating a broad class of classical and quantum Hamiltonians in lattices with arbitrary topologies. The device contains only a single actively controlled optical component -- a phase-modulated MZI -- and can be reprogrammed to emulate a wide variety of systems, such as chiral states in a Hall ladder, synthetic gauge potentials, and high-dimensional lattices. Our proposed device is experimentally appealing and opens new possibilities for studying fundamental topological and many-body physics.

\vspace{5mm} 

\emph{Acknowledgements} ---
This work was supported by a Vannevar Bush Faculty Fellowship from the U.S. Department of Defense (N00014-17-1-3030) and by an AFOSR MURI project (Grant No. FA9550-22-1-0339).  A.D. was partially supported by a National Quantum Lab at Maryland (Q-Lab) seed grant.

%


\end{document}


\title{Supplementary Materials for \\
``Programmable device for quantum simulation in arbitrary topologies''}

\author{Ben Bartlett}
\affiliation{Department of Applied Physics, Stanford University, Stanford, California 94305, USA}
\affiliation{Department of Electrical Engineering, Stanford University, Stanford, California 94305, USA}

\author{Olivia Y. Long}
\affiliation{Department of Applied Physics, Stanford University, Stanford, California 94305, USA}

\author{Avik Dutt}
\affiliation{Department of Mechanical Engineering, University of Maryland, College Park, Maryland 20742, USA}
\affiliation{Institute for Physical Science and Technology, University of Maryland, College Park, Maryland 20742, USA}

\author{Shanhui Fan}
\email{shanhui@stanford.edu}
\affiliation{Department of Electrical Engineering, Stanford University, Stanford, California 94305, USA}

\maketitle

In this Supplementary Materials document, we give more detailed presentations of the results described the main text. In Sections \ref{sec:hopping_coefficients} and \ref{sec:onsite_interaction_potential} we derive a correspondence between the device physics and the terms in the target Hamiltonian, and in Section \ref{sec:simulation_details} we provide additional details on the simulation methods used to tractably simulate interacting bosons on large lattices.

\section{Deriving hopping coefficients between sites}
\label{sec:hopping_coefficients}

The Hamiltonians of interest take the form:
\begin{equation}
\hat{H} = 
\sum_{\langle m, n \rangle} \left( \kappa_{mn} e^{i \alpha_{mn}} \hat{a}^\dagger_m \hat{a}_n + \mathrm{H.c.} \right)
+ \mu \sum_m \hat{a}^\dagger_m \hat{a}_m 
+ U \sum_{m} \hat{a}^\dagger_m \hat{a}^\dagger_m \hat{a}_m \hat{a}_m,
\label{eq:hamiltonian_em_sm}
\end{equation}
where $\hat{a}^\dagger_{m}$ creates a boson at node $m$, $\kappa_{m,n}$ denotes the hopping coefficients between sites $m$ and $n$, with the Hermitian requirement that $\kappa_{m,n} = \kappa_{n,m}$, $\mu$ is the single-body energy per site, $U$ is the Hubbard interaction strength, and $\alpha_{m,n}$ is a phase shift accumulated by moving from site $m$ to $n$. The two-particle summation in the first term is taken over all connected sites $\langle m,n \rangle$, where the connectivity of the simulated system is determined with suitable choice of $\left\{\kappa_{m,n}\right\}$.

A system evolving under the Hamiltonian $\hat{H}$ in Eq. \ref{eq:hamiltonian_em_sm} will have a propagator (with $\hbar=1$) of the form:

\begin{equation}
e^{-i\hat{H}t} = \exp \left[ -it \left(-\sum_{\langle m, n \rangle} \kappa_{mn} e^{i \alpha_{mn}}  \hat{a}^\dagger_m \hat{a}_n + \mu \sum_m \hat{a}^\dagger_m \hat{a}_m + U \sum_{m} \hat{a}^\dagger_m \hat{a}^\dagger_m \hat{a}_m \hat{a}_m \right)\right]
\end{equation}

If we add Hermitian constraints to the Hamiltonian, we have that $\kappa_{mn} = \kappa_{nm}$ and $\alpha_{mn} = -\alpha_{nm}$, so we can write:
\begin{equation}
e^{-i\hat{H}t} 
= \exp \left[ -it \left(-\sum_{\langle m, n \rangle} \kappa_{mn} \left( e^{i \alpha_{mn}}  \hat{a}^\dagger_m \hat{a}_n  + e^{-i \alpha_{mn}} \hat{a}^\dagger_n \hat{a}_m \right) + \mu \sum_m \hat{a}^\dagger_m \hat{a}_m + U \sum_{m} \hat{a}^\dagger_m \hat{a}^\dagger_m \hat{a}_m \hat{a}_m \right)\right],
\end{equation}
where the product over $\langle m, n \rangle$ now implicitly avoids double-counting, as we have explicitly included the Hermitian conjugate in the first term. We can series expand this as $e^{t(X+Y)} = e^{tX} e^{tY} e^{-\frac{t^2}{2} [X,Y]} e^{\frac{t^3}{6}(2[Y,[X,Y]]+ [X,[X,Y]] )} \cdots$ (the Suzuki-Trotter expansion) to separate the summation into a product of matrix exponentials:
\begin{align}
\begin{split}
e^{-i\hat{H}t} 
&\approx \exp\left(i t\sum_{\langle m, n \rangle} \kappa_{mn} \left( e^{i \alpha_{mn}}  \hat{a}^\dagger_m \hat{a}_n  + e^{-i \alpha_{mn}} \hat{a}^\dagger_n \hat{a}_m \right) \right) \times e^{-it \left( \mu \sum_m \hat{a}^\dagger_m \hat{a}_m + U \sum_m \hat{a}^\dagger_m \hat{a}_m \hat{a}_m \hat{a}^\dagger_m \right)} \\ 
&\qquad \times \exp \left( -\frac{t^2}{2} \left[\sum_{\langle m,n\rangle}i \kappa_{mn} \left( e^{i \alpha_{mn}}  \hat{a}^\dagger_m \hat{a}_n  + e^{-i \alpha_{mn}} \hat{a}^\dagger_n \hat{a}_m \right),\quad\sum_{m}\mu\adag_{m}\ahat_{m} + \sum_{m}U\adag_{m}\adag_{m}\ahat_{m}\ahat_{m}\right] \right) \\ 
&\equiv \exp\left(i t\sum_{\langle m, n \rangle} i \kappa_{mn} \left( e^{i \alpha_{mn}}  \hat{a}^\dagger_m \hat{a}_n  + e^{-i \alpha_{mn}} \hat{a}^\dagger_n \hat{a}_m \right) \right) \times e^{-it \left( \mu \sum_m \hat{a}^\dagger_m \hat{a}_m + U \sum_m \hat{a}^\dagger_m \hat{a}_m \hat{a}_m \hat{a}^\dagger_m \right)} \times \varepsilon_{\kappa,\mu} \times \varepsilon_{\kappa,U}
\end{split}
\label{eq:first_commutator_split}
\end{align}
We evaluate the commutator error terms $\varepsilon_{\mu}$ and $\varepsilon_{U}$ in Section \ref{sec:commutator_bullshit} and find that $\varepsilon_{\mu} = \id$ since the $\kappa_{mn}$ and $\mu$ terms commute, and $\varepsilon_{U} = \exp\left( -2 t^2 U\sum_{\langle m,n\rangle}\kappa_{mn}\cos\alpha_{mn}\left((\adag_{m}\ahat_{m})\adag_{n}\ahat_{m}-\adag_{m}\ahat_{n}(\adag_{m}\ahat_{m})\right) \right)$.

We now perform a second series expansion on the first term to separate the exponential of sum into a product of exponentials:
\begin{align}
\begin{split}
e^{-i\hat{H}t} 
&\approx \left(\prod_{\langle m, n \rangle} \exp i \kappa_{mn} \left( e^{i \alpha_{mn}}  \hat{a}^\dagger_m \hat{a}_n  + e^{-i \alpha_{mn}} \hat{a}^\dagger_n \hat{a}_m \right) 
\right)^t \times e^{-it \left( \sum_m \mu \hat{a}^\dagger_m \hat{a}_m + U \hat{a}^\dagger_m \hat{a}^\dagger_m \hat{a}_m \hat{a}_m \right)} \times \varepsilon_{U} \\
&\qquad \times \exp \left(- \frac{t^2}{2} \sum_{\langle j, k \ne k'\rangle} [\kappa_{jk} (e^{i \alpha_{jk}} \adag_j \hat{a}_k + e^{-i \alpha_jk} \adag_k \hat{a}_j), \; \kappa_{jk'} (e^{i \alpha_{jk'}} \adag_j \hat{a}_{k'} + e^{-i \alpha_{jk'}} \adag_{k'} \hat{a}_j)]\right) \\ 
&\equiv \left(\prod_{\langle m, n \rangle} \exp i \kappa_{mn} \left( e^{i \alpha_{mn}}  \hat{a}^\dagger_m \hat{a}_n  + e^{-i \alpha_{mn}} \hat{a}^\dagger_n \hat{a}_m \right) 
\right)^t \times e^{-it \left( \sum_m \mu \hat{a}^\dagger_m \hat{a}_m + U \hat{a}^\dagger_m \hat{a}^\dagger_m \hat{a}_m \hat{a}_m \right)} \times \varepsilon_{U} \times \varepsilon_{\kappa}.
\end{split}
\end{align}
We can expand the exponentials of the error terms as $e^A \approx \id + A$ to obtain error scaling on the final result:
\begin{align}
\begin{split}
e^{-i\hat{H}t} &= \left(\prod_{\langle m, n \rangle} \exp\left(i \kappa_{mn} e^{i \alpha_{mn}}  \hat{a}^\dagger_m \hat{a}_n + \text{H.c.} \right) 
\right)^t \times e^{-it \left( \sum_m \mu \hat{a}^\dagger_m \hat{a}_m + U \hat{a}^\dagger_m \hat{a}^\dagger_m \hat{a}_m \hat{a}_m \right)} \\
&\qquad \times \left(\id + \mathcal{O}(\kappa U \cos \alpha) \sum_{\langle m,n \rangle} \left(\adag_{n}\adag_{m}\ahat_{m}\ahat_{m}-\ahat_{n}\adag_{m}\adag_{m}\ahat_{m}\right) + \mathcal{O}(\kappa^2) \sum_{\langle j,k\ne k' \rangle} [\adag_j \hat{a}_k, \hat{a}_j \adag_{k'}] \right)^{-t^2/2}
\end{split}
\end{align}
where $\kappa, \alpha$ is shorthand for typical (or uniform) values of $\kappa_{mn}, \alpha_{mn}$. If we have small coefficients $\kappa, U \ll 1$, then $\mathcal{O}(\kappa^2 + \kappa U)$ is negligible, and we ignore the commutator error term going forward. If $\kappa, U$ is not small, we can reduce the emulated values of $\kappa_{mn}$, $\mu$, and $U$ by some constant factor $C$ and run the emulation for a commensurately longer wall clock time $Ct$. Thus, to within $\mathcal{O}(\kappa^2 + \kappa U)$, we can write the propagator as:
\begin{equation}
e^{-i\hat{H}t} = \left[\prod_{\langle m, n \rangle} \exp\left(i \kappa_{mn} e^{i \alpha_{mn}}  \hat{a}^\dagger_m \hat{a}_n + \text{H.c.}\right) 
\right]^t \times e^{-it \left( \mu \sum_m \hat{a}^\dagger_m \hat{a}_m + U \sum_m \hat{a}^\dagger_m \hat{a}^\dagger_m \hat{a}_m \hat{a}_m \right)}.
\label{eq:propagator_sm}
\end{equation}

Now consider the tunable MZI connecting the storage ring to the register ring as shown in Figure 1 of the main text. Define bosonic operators $\adag_1$ and $\adag_2$ which create right-moving photons at the input waveguides to the MZI and operators $\bdag_1$ and $\bdag_2$ which create right-moving output photons. The transfer matrix of the MZI if the internal phase shifter is set to an angle $\theta$ and the external phase shifters are set to $\pm \phi$ is $T = R_z(-\phi) H R_z(\theta) H R_z(+\phi)$. Thus, we can relate the output modes to the input modes as:
\begin{equation}
\begin{bmatrix} 
\bdag_1 \\ 
\bdag_2
\end{bmatrix} 
= 
\begin{pmatrix}
\cos \frac{\theta}{2} & i e^{i \phi} \sin \frac{\theta}{2} \\
i e^{-i\phi} \sin \frac{\theta}{2} & \cos \frac{\theta}{2}
\end{pmatrix}
\begin{bmatrix} 
\adag_1 \\ 
\adag_2
\end{bmatrix}
= \exp\left[ i \frac{\theta}{2} \left(e^{i\phi} \adag_1 \hat{a}_2 + e^{-i\phi} \adag_2 \hat{a}_1\right) \right]
\begin{bmatrix} 
\adag_1 \\ 
\adag_2
\end{bmatrix}.
\end{equation}
If we define $\kappa \equiv \frac{\theta}{2}$ and $\alpha \equiv \phi$, then we obtain the desired transfer matrix from which we can construct the first part of the propagator in Eq. \ref{eq:propagator_sm}:
\begin{equation}
T_{12} = \exp\left[ i \kappa \left(e^{i\alpha} \adag_1 \hat{a}_2 + e^{-i\alpha} \adag_2 \hat{a}_1\right) \right].
\end{equation}

If we apply a sequence of passes of the photon pulses through the MZI, then by appropriately choosing values of $\theta,\phi$ to match $\kappa_{mn}, \alpha_{mn}$, we obtain a total transfer matrix of:
\begin{equation}
T_{\langle m, n \rangle} = \prod_{\langle m, n \rangle} \exp\left( i \kappa_{mn} \left(e^{i\alpha_{mn}} \adag_m \hat{a}_n + e^{-i\alpha_{mn}} \adag_n \hat{a}_m\right) \right),
\end{equation}
so after $t$ repetitions of this sequence of passes, the total accumulated operation is:
\begin{equation}
T_{\langle m, n \rangle} = \left[\prod_{\langle m, n \rangle} \exp\left( i \kappa_{mn} \left(e^{i\alpha_{mn}} \adag_m \hat{a}_n + e^{-i\alpha_{mn}} \adag_n \hat{a}_m\right) \right)\right]^t,
\end{equation}
which is exactly the desired form from the propagator in Eq. \ref{eq:propagator_sm}.

\subsection{Evaluating the commutator error terms}
\label{sec:commutator_bullshit}

We can split the commutator error term in the last line of Eq. \ref{eq:first_commutator_split} into two parts: $\exp\left( -it^2 /2 (\varepsilon_{\kappa, \mu} + \varepsilon_{\kappa, U})\right)$. Starting with the $\kappa_{mn}, \mu$ commutator, we have:
\begin{align}
\begin{split}
\varepsilon_{\kappa, \mu}	&=\left[\sum_{\langle m,n\rangle}\kappa_{mn}\p{e^{i\alpha_{mn}}\adag_{m}\ahat_{n}+e^{-i\alpha_{mn}}\adag_{n}\ahat_{m}},\quad\sum_{m}\mu\adag_{m}\ahat_{m}\right] \\
&=\sum_{\langle m,n\rangle}\kappa_{mn}\p{e^{i\alpha_{mn}}\adag_{m}\ahat_{n}+e^{-i\alpha_{mn}}\adag_{n}\ahat_{m}}\sum_{o}\mu\adag_{o}\ahat_{o}-\sum_{o}\mu\adag_{o}\ahat_{o}\sum_{\langle m,n\rangle}\kappa_{mn}\p{e^{i\alpha_{mn}}\adag_{m}\ahat_{n}+e^{-i\alpha_{mn}}\adag_{n}\ahat_{m}} \\
&=\sum_{o}\mu\sum_{\langle m,n\rangle}\kappa_{mn}\left(\p{e^{i\alpha_{mn}}\adag_{m}\ahat_{n}+e^{-i\alpha_{mn}}\adag_{n}\ahat_{m}}\adag_{o}\ahat_{o}-\adag_{o}\ahat_{o}\p{e^{i\alpha_{mn}}\adag_{m}\ahat_{n}+e^{-i\alpha_{mn}}\adag_{n}\ahat_{m}}\right).
\end{split}
\end{align}
Imposing $\delta_{om}$ and $\delta_{on}$, for each point $o$, we have $\varepsilon_{mu}=\varepsilon_{o=m}+\varepsilon_{o=n}$:
\begin{align}
\begin{split}
\varepsilon_{o=m}	&=\mu\sum_{\langle m,n\rangle}\kappa_{mn}\left(\p{e^{i\alpha_{mn}}\adag_{m}\ahat_{n}+e^{-i\alpha_{mn}}\adag_{n}\ahat_{m}}\adag_{m}\ahat_{m}-\adag_{m}\ahat_{m}\p{e^{i\alpha_{mn}}\adag_{m}\ahat_{n}+e^{-i\alpha_{mn}}\adag_{n}\ahat_{m}}\right) \\
\varepsilon_{o=n}	&=\mu\sum_{\langle m,n\rangle}\kappa_{mn}\left(\p{e^{i\alpha_{mn}}\adag_{m}\ahat_{n}+e^{-i\alpha_{mn}}\adag_{n}\ahat_{m}}\adag_{n}\ahat_{n}-\adag_{n}\ahat_{n}\p{e^{i\alpha_{mn}}\adag_{m}\ahat_{n}+e^{-i\alpha_{mn}}\adag_{n}\ahat_{m}}\right).
\end{split}
\end{align}
Using that $\ahat\adag\ahat = \adag\ahat\ahat+\ahat$ and $\adag\adag\ahat=\ahat\adag\adag-\adag$, we expand these expressions to obtain:
\begin{align}
\begin{split}
\varepsilon_{o=m}&=\mu\sum_{\langle m,n\rangle}\kappa_{mn}\left(e^{i\alpha_{mn}}\adag_{m}\ahat_{n}\p{\adag_{m}\ahat_{m}}+e^{-i\alpha_{mn}}\adag_{n}\ahat_{m}\p{\adag_{m}\ahat_{m}}-e^{i\alpha_{mn}}\p{\adag_{m}\ahat_{m}}\adag_{m}\ahat_{n}-e^{-i\alpha_{mn}}\p{\adag_{m}\ahat_{m}}\adag_{n}\ahat_{m}\right)\\
&=\mu\sum_{\langle m,n\rangle}\kappa_{mn}\left(e^{i\alpha_{mn}}\ahat_{n}\adag_{m}\adag_{m}\ahat_{m}+e^{-i\alpha_{mn}}\adag_{n}\ahat_{m}\adag_{m}\ahat_{m}-e^{i\alpha_{mn}}\ahat_{n}\adag_{m}\ahat_{m}\adag_{m}-e^{-i\alpha_{mn}}\adag_{n}\adag_{m}\ahat_{m}\ahat_{m}\right)\\
&=\mu\sum_{\langle m,n\rangle}\kappa_{mn}\left(e^{i\alpha_{mn}}\ahat_{n}\adag_{m}\adag_{m}\ahat_{m}+e^{-i\alpha_{mn}}\adag_{n}\p{\adag_{m}\ahat_{m}+1}\ahat_{m}-e^{i\alpha_{mn}}\ahat_{n}\adag_{m}\p{\adag_{m}\ahat_{m}+1}-e^{-i\alpha_{mn}}\adag_{n}\adag_{m}\ahat_{m}\ahat_{m}\right)\\
&=\mu\sum_{\langle m,n\rangle}\kappa_{mn}\left(e^{-i\alpha_{mn}}\adag_{n}\ahat_{m}-e^{i\alpha_{mn}}\ahat_{n}\adag_{m}\right) \\ 
\varepsilon_{o=n}&=\mu\sum_{\langle m,n\rangle}\kappa_{mn}\left(e^{i\alpha_{mn}}\adag_{m}\ahat_{n}\p{\adag_{n}\ahat_{n}}+e^{-i\alpha_{mn}}\adag_{n}\ahat_{m}\p{\adag_{n}\ahat_{n}}-e^{i\alpha_{mn}}\p{\adag_{n}\ahat_{n}}\adag_{m}\ahat_{n}-e^{-i\alpha_{mn}}\p{\adag_{n}\ahat_{n}}\adag_{n}\ahat_{m}\right) \\ 
&=\mu\sum_{\langle m,n\rangle}\kappa_{mn}\left(e^{i\alpha_{mn}}\adag_{m}\ahat_{n}-e^{-i\alpha_{mn}}\ahat_{m}\adag_{n}\right).
\end{split}
\end{align}
Therefore $\varepsilon_{\kappa,\mu}=\varepsilon_{o=m}+\varepsilon_{o=n}=0$, so the $\kappa_{mn}$ and $\mu$ terms commute. Now we evaluate the $\varepsilon_{\kappa,U}$ error term. Starting as before we have:
\begin{align}
\begin{split}
\varepsilon_{U}&=\left[\sum_{\langle m,n\rangle}\kappa_{mn}\p{e^{i\alpha_{mn}}\adag_{m}\ahat_{n}+e^{-i\alpha_{mn}}\adag_{n}\ahat_{m}},\quad\sum_{m}U\adag_{m}\adag_{m}\ahat_{m}\ahat_{m}\right]\\&=\sum_{\langle m,n\rangle}\kappa_{mn}\p{e^{i\alpha_{mn}}\adag_{m}\ahat_{n}+e^{-i\alpha_{mn}}\adag_{n}\ahat_{m}}\sum_{o}U\adag_{o}\adag_{o}\ahat_{o}\ahat_{o}-\sum_{o}U\adag_{o}\adag_{o}\ahat_{o}\ahat_{o}\sum_{\langle m,n\rangle}\kappa_{mn}\p{e^{i\alpha_{mn}}\adag_{m}\ahat_{n}+e^{-i\alpha_{mn}}\adag_{n}\ahat_{m}}\\&=\sum_{o}U\sum_{\langle m,n\rangle}\kappa_{mn}\left(\p{e^{i\alpha_{mn}}\adag_{m}\ahat_{n}+e^{-i\alpha_{mn}}\adag_{n}\ahat_{m}}\adag_{o}\adag_{o}\ahat_{o}\ahat_{o}-\adag_{o}\adag_{o}\ahat_{o}\ahat_{o}\p{e^{i\alpha_{mn}}\adag_{m}\ahat_{n}+e^{-i\alpha_{mn}}\adag_{n}\ahat_{m}}\right).
\end{split}
\end{align}
Once again we impose $\delta_{om}, \delta_{on}$, so for each point $o$ we have $\varepsilon_{\kappa,U}=\varepsilon_{o=m}+\varepsilon_{o=n}$:
\begin{align}
\begin{split}
\varepsilon_{o=m}&=U\sum_{\langle m,n\rangle}\kappa_{mn}\left(\p{e^{i\alpha_{mn}}\adag_{m}\ahat_{n}+e^{-i\alpha_{mn}}\adag_{n}\ahat_{m}}\adag_{m}\adag_{m}\ahat_{m}\ahat_{m}-\adag_{m}\adag_{m}\ahat_{m}\ahat_{m}\p{e^{i\alpha_{mn}}\adag_{m}\ahat_{n}+e^{-i\alpha_{mn}}\adag_{n}\ahat_{m}}\right)\\\varepsilon_{o=n}&=U\sum_{\langle m,n\rangle}\kappa_{mn}\left(\p{e^{i\alpha_{mn}}\adag_{m}\ahat_{n}+e^{-i\alpha_{mn}}\adag_{n}\ahat_{m}}\adag_{n}\adag_{n}\ahat_{n}\ahat_{n}-\adag_{n}\adag_{n}\ahat_{n}\ahat_{n}\p{e^{i\alpha_{mn}}\adag_{m}\ahat_{n}+e^{-i\alpha_{mn}}\adag_{n}\ahat_{m}}\right).
\end{split}
\end{align}
Using that $\ahat\adag\adag\ahat\ahat=\adag\adag\ahat\ahat\ahat+2\adag\ahat\ahat$ and $\ahat^{\dag}\adag\adag\ahat\ahat=\adag\adag\ahat\ahat\adag-2\adag\adag\ahat$, we obtain:
\begin{align}
\begin{split}
\varepsilon_{o=m}&=U\sum_{\langle m,n\rangle}\kappa_{mn}
\begin{multlined}[t]
(e^{i\alpha_{mn}}\adag_{m}\ahat_{n}\p{\adag_{m}\adag_{m}\ahat_{m}\ahat_{m}}+e^{-i\alpha_{mn}}\adag_{n}\ahat_{m}\p{\adag_{m}\adag_{m}\ahat_{m}\ahat_{m}} \\ 
-e^{i\alpha_{mn}}\p{\adag_{m}\adag_{m}\ahat_{m}\ahat_{m}}\adag_{m}\ahat_{n}-e^{-i\alpha_{mn}}\p{\adag_{m}\adag_{m}\ahat_{m}\ahat_{m}}\adag_{n}\ahat_{m})
\end{multlined}\\
&=U\sum_{\langle m,n\rangle}\kappa_{mn}
\begin{multlined}[t]
(e^{i\alpha_{mn}}\adag_{m}\ahat_{n}\p{\adag_{m}\adag_{m}\ahat_{m}\ahat_{m}}+e^{-i\alpha_{mn}}\adag_{n}\ahat_{m}\p{\adag_{m}\adag_{m}\ahat_{m}\ahat_{m}} \\
-e^{i\alpha_{mn}}\adag_{m}\ahat_{n}\p{\adag_{m}\adag_{m}\ahat_{m}\ahat_{m}}-e^{-i\alpha_{mn}}\adag_{n}\ahat_{m}\p{\adag_{m}\adag_{m}\ahat_{m}\ahat_{m}} \\
-2e^{i\alpha_{mn}}\ahat_{n}\adag_{m}\adag_{m}\ahat_{m}+2e^{-i\alpha_{mn}}\adag_{n}\adag_{m}\ahat_{m}\ahat_{m})
\end{multlined}\\
&=U\sum_{\langle m,n\rangle}2\kappa_{mn}\left(e^{-i\alpha_{mn}}\p{\adag_{m}\ahat_{m}}\adag_{n}\ahat_{m}-e^{+i\alpha_{mn}}\adag_{m}\ahat_{n}\p{\adag_{m}\ahat_{m}}\right)\\
\varepsilon_{o=n}&=U\sum_{\langle m,n\rangle}2\kappa_{mn}\left(e^{+i\alpha_{mn}}\p{\adag_{n}\ahat_{n}}\adag_{m}\ahat_{n}-e^{-i\alpha_{mn}}\adag_{n}\ahat_{m}\p{\adag_{n}\ahat_{n}}\right).
\end{split}
\end{align}
Therefore the error term $\varepsilon_{\kappa,U}=\varepsilon_{o=m}+\varepsilon_{o=n}$ is:
\begin{align}
\begin{split}
\varepsilon_{\kappa,U}&=U\sum_{\langle m,n\rangle}2\kappa_{mn}\left(e^{-i\alpha_{mn}}\p{\adag_{m}\ahat_{m}}\adag_{n}\ahat_{m}-e^{+i\alpha_{mn}}\adag_{m}\ahat_{n}\p{\adag_{m}\ahat_{m}}\right)\\
&+U\sum_{\langle m,n\rangle}2\kappa_{mn}\left(e^{+i\alpha_{mn}}\p{\adag_{n}\ahat_{n}}\adag_{m}\ahat_{n}-e^{-i\alpha_{mn}}\adag_{n}\ahat_{m}\p{\adag_{n}\ahat_{n}}\right).
\end{split}
\end{align}
In the second term, we will swap the m and n indices. Since the second summation is separate from the first and this is just a notational change, we do not switch $\alpha_{mn}=-\alpha_{nm}$, so we have:
\begin{align}
\begin{split}
\varepsilon_{\kappa,U}&=U\sum_{\langle m,n\rangle}2\kappa_{mn}\left(e^{-i\alpha_{mn}}\p{\adag_{m}\ahat_{m}}\adag_{n}\ahat_{m}-e^{+i\alpha_{mn}}\adag_{m}\ahat_{n}\p{\adag_{m}\ahat_{m}}\right)\\
&+U\sum_{\langle m,n\rangle}2\kappa_{mn}\left(e^{+i\alpha_{mn}}\p{\adag_{m}\ahat_{m}}\adag_{n}\ahat_{m}-e^{-i\alpha_{mn}}\adag_{m}\ahat_{n}\p{\adag_{m}\ahat_{m}}\right).
\end{split}
\end{align}
Combining these expressions, we obtain the final result:
\begin{align}
\begin{split}
\varepsilon_{\kappa,U}&=U\sum_{\langle m,n\rangle}2\kappa_{mn}\left(\p{e^{-i\alpha_{mn}}+e^{+i\alpha_{mn}}}\p{\adag_{m}\ahat_{m}}\adag_{n}\ahat_{m}-\p{e^{+i\alpha_{mn}}+e^{-i\alpha_{mn}}}\adag_{m}\ahat_{n}\p{\adag_{m}\ahat_{m}}\right)\\
&=4U\sum_{\langle m,n\rangle}\kappa_{mn}\cos\alpha_{mn}\left(\p{\adag_{m}\ahat_{m}}\adag_{n}\ahat_{m}-\adag_{m}\ahat_{n}\p{\adag_{m}\ahat_{m}}\right).
\end{split}
\end{align}
Thus, employing the discrete time evolution for the emulation of the $\kappa_{mn}$ terms with the MZI does impose an error term scaling as $\mathcal{O}\left(U \kappa \cos \alpha\right)$. However, for $\alpha_{mn}=\pi/2$, this error term vanishes and perfect emulation is possible. This is interesting and particularly fortunate as $\alpha=\pi/2$ is the hopping phase at which most topological effects are maximized in 2D.

\section{Deriving nonlinear on-site interaction potential}
\label{sec:onsite_interaction_potential}

For lossless and dispersionless media, the polarization density can be expanded in the form \cite{boyd_nonlinear_optics_book}
\begin{equation}
P(t) = \epsilon_0 \left[\chi^{(1)} E(t) + \chi^{(2)}E^2(t) + \chi^{(3)}E^3(t) + \cdots \right],
\end{equation}
where $\epsilon_0$ is the permittivity of free space, $\chi^{(n)}$ terms are optical susceptibilities, and where where $P(t)$ and $E(t)$ are expressed as scalar quantities for simplicity. For media with inversion symmetry (e.g. glass), the $\chi^{(2)}$ term vanishes. \cite{bloembergen_nonlinear_optics_book}

We can write $E(t)$ in the frequency domain as a summation is over frequencies $\omega_n$: 
\begin{equation}
    E(t) = \sum_n \left[ E(\omega_n)e^{-i\omega_n t} + E(-\omega_n)e^{+i\omega_n t}\right],
\end{equation}
where $E(\omega_n) = \lvert E_n \rvert e^{i k \cdot r}/2 = E(-\omega_n)^*$.
The polarization density can similarly be expressed as 
\begin{equation}
    P(t) = \sum_n [ P (\omega_n)e^{-i\omega_n t} + P(-\omega_n)e^{+i\omega_n t} ].
\end{equation}

Since the Kerr nonlinearity arises from the third-order susceptibility $\chi^{(3)}$, we will focus on this term and refer to it as $\mathcal{P}_\mathrm{NL}(\omega)$. We can write this nonlinear polarization vector as:
\begin{equation}
\mathcal{P}_{\mathrm{NL},i}(\omega) = \epsilon_0 \sum_{ijkl} \sum_{(mno)} \chi^{(3)}_{ijkl}(\omega_n, \omega_m, \omega_o) \times E_j(\omega_m)E_k(\omega_n)E_l(\omega_o),
\end{equation}
where $\chi^{(3)}$ is a rank-4 tensor, $i, j, k, l$ are Cartesian coordinates, and $(mno)$ refers to the number of permutations of the distinguishable $E$ fields that yield the same $\omega_m + \omega_n + \omega_o$ term, following the notation of Ref. \citenum{boyd_nonlinear_optics_book}. The term that gives rise to the intensity-dependent refractive index is:
\begin{equation}
\mathcal{P}_{\mathrm{NL},i}(\omega) = 3\epsilon_0 \sum_{ijkl} \chi^{(3)}_{ijkl}(\omega) \times E_j(\omega)E_k(\omega)E_l(-\omega).
\end{equation}

Assuming the medium is isotropic and homogeneous (e.g. glass), the susceptibility becomes polarization-independent and thus can be treated as a scalar. \cite{Strekalov_2016} Without loss of generality, assume the input $E$ fields are linearly-polarized in the $x$ direction, denoted by index $j, k, l = 1$. The electric field energy density for the third-order nonlinear polarization can be expressed as \cite{boyd_nonlinear_optics_book}: 
\begin{align}
\mathcal{U}_\mathrm{NL} &= \frac{3\epsilon_0}{4} \sum_{i111}  \chi_{i111}^{(3)} \times E_i^*(\omega + \omega - \omega)E_1(\omega)E_1(\omega)E_1^*(\omega) \\
&= \frac{3\epsilon_0}{4} \sum_{i111}  \chi_{i111}^{(3)} \times E_i^*(\omega + \omega - \omega)|E_1(\omega)|^2 E_1(\omega)
\end{align}

From this expression, the nonlinear polarization $\mathcal{P}_{\mathrm{NL}}$ along a direction $i$ can be obtained by differentiating with respect to $E_i^*$:
\begin{equation}
    \mathcal{P}_{\mathrm{NL},i} = \frac{\partial \mathcal{U}_{\mathrm{NL}}}{\partial E^*_i}
\end{equation}

Defining $z$ to be the direction of field propagation in the fiber, the electric fields in the ring fiber can now be quantized as follows: \cite{loudon_quantum_optics_book, avik_exp_band_struc_nature_2019} 

\begin{equation}\label{quantized_E}
E = \sum_{k} i \left ( \frac{\hbar \omega_k}{2 \epsilon_0 n_0^2 V} \right )^{1/2} (\hat{a}_{k} e^{-i (\omega_k t - k z)}+ \hat{a}^{\dagger}_{k} e^{+i (\omega_k t - k z) })
\end{equation}


where $\hat{a}^{\dagger}_{k}$ and $\hat{a}_{k}$ are the bosonic creation and annihilation operators at the frequency $\omega_k$. $E$ has been normalized to the volume $V$ of the fiber, and $n_0$ is the refractive index of the fiber. 
We can express the operators in real space through Fourier transform \cite{Shen2009TheoryAtom}: 
\begin{align}
    \hat{a}_{k} &= \int_{0}^{L} \hat{a}_{\omega_k} (z) e^{-ikz} dz \\
    \hat{a}^{\dagger}_{k} &= \int_{0}^{L} \hat{a}^{\dagger}_{\omega_k} (z) e^{+i k z} dz,
\end{align}
where $\hat{a}_{\omega_k} (z)$ and $\hat{a}^\dagger_{\omega_k} (z)$ annihilate and create a photon with frequency $\omega_{k}$ at position $z$ in real space, respectively, and where $L$ is the length of the fiber.
Since we employ monochromatic light at frequency $\omega_0$, we only have modes within a narrow bandwidth around $\omega_0$ (${k}_0$). Thus, we can replace the $k$-space operators in Equation \ref{quantized_E} with the real-space operators $\hat{a}_{\omega_0} (z)$ and $\hat{a}^{\dagger}_{\omega_0} (z)$, which we will hereafter refer to as $a$ and $a^{\dagger}$, respectively.
Substituting the quantized electric fields into our expression for the energy density and applying the rotating-wave approximation, we have the following terms in our expression for the total energy of the third-order nonlinear susceptibility: 
\begin{align}
\begin{split}
H = \int_V \mathcal{U}_\mathrm{NL} d^3\mathbf{r} &= \frac{3\epsilon_0}{4} \int_V \chi^{(3)} |E|^4 d^3\mathbf{r} \\
&= \frac{3\epsilon_0 \hbar^2 \omega_0^2}{16 \epsilon_0^2 n_0^4 V^2} \int_V \chi^{(3)} (\hat{a} \hat{a}^{\dagger} + \hat{a} \hat{a} e^{-2i(\omega_0 t - k_0 z)} +\hat{a}^{\dagger} \hat{a}^{\dagger} e^{+2i(\omega_0 t - k_0 z)} + \hat{a}^{\dagger} \hat{a} )^2 d^3\mathbf{r} \\
&\approx \frac{3\hbar^2 \omega_0^2}{16 \epsilon_0 n_0^4 V^2} \int_V \chi^{(3)} (6 \hat{a}^{\dagger} \hat{a}^{\dagger} \hat{a} \hat{a} + 12 \hat{a}^{\dagger} \hat{a} + 3 ) d^3\mathbf{r} \\
&= \left( \frac{9 \hbar^2 \omega_0^2}{8 \epsilon_0 n_0^4 V^2} \int_V \chi^{(3)}  d^3\mathbf{r} \right ) \hat{a}^{\dagger} \hat{a}^{\dagger} \hat{a} \hat{a} + C,
\end{split}
\end{align}
where $C$ is some constant corresponding to an overall energy shift \cite{local_Hamilt_synth_dim_luqi_20}.
This gives the desired $\hat{a}^\dagger a^{\dagger} \hat{a} \hat{a}$ term in the Hamiltonian, where the coefficient $U = \frac{9 \hbar^2 \omega_0^2}{8 \epsilon_0 n_0^4 V^2} \int_V \chi^{(3)} d^3\mathbf{r}$. Note that we only retain the two-photon interaction term $\hat{a}^\dagger_m \hat{a}^\dagger_n \hat{a}_p \hat{a}_q$ where $m=n=p=q$ in the Hamiltonian since only spatially localized photons in the same time bin interact to achieve the optical Kerr effect, and we are assuming monochromatic light with one frequency mode $\omega_0$.

\section{Simulation details}
\label{sec:simulation_details}

Here we outline the details of the simulations we use to compare the evolution of the states under the emulated Hamiltonians against the exact target Hamiltonians. We use a custom simulation method built with \texttt{QuTiP} \cite{Johansson2012QuTiP:Systems} for all simulations presented in this paper. 

\subsection{Tractable simulation with a Ponomarev state representation}

\begin{figure}[b]
\centering 
\includegraphics[width=.85\textwidth]{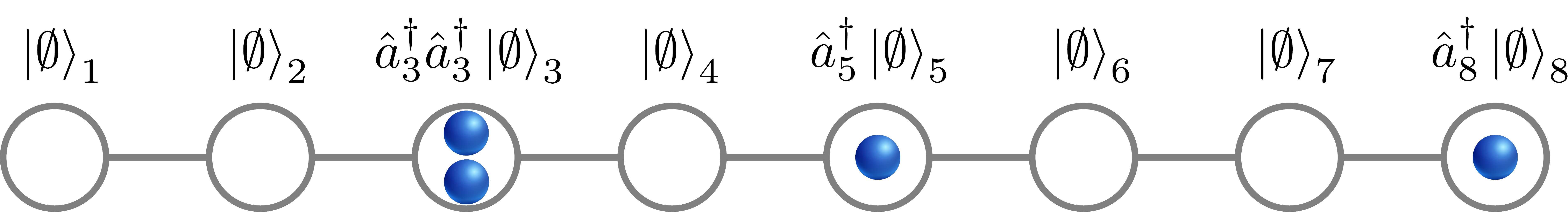}
\caption{
An example Fock state $\ket{\psi} = \ket{0,0,2,0,1,0,0,1}$, which can be re-indexed by boson number as $\ket{\psi'} = \ket{8,5,3,3}.$ This maps to the Ponomarev basis state $\ket{n_\psi} = \ket{112}$.
}
\label{fig:fock_to_ponomarev}
\end{figure}

\begin{figure}[t]
\centering 
\includegraphics[width=.9\textwidth]{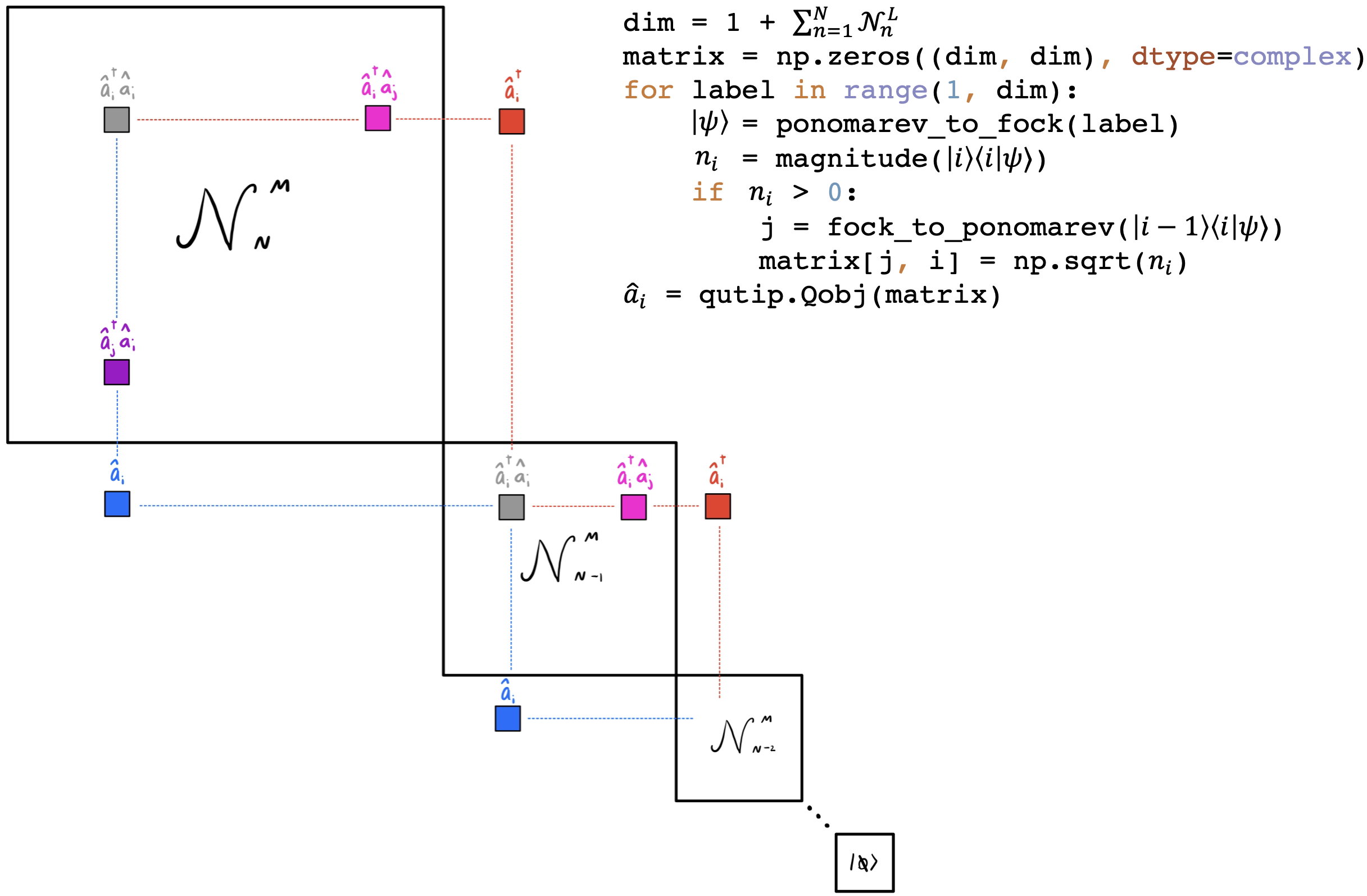}
\caption{
Construction of an annihilation operator in the bosonic lattice representation. Python pseudocode showing the process for constructing a \texttt{QuTiP} object from the Ponomarev indexing functions is included in the upper right.
}
\label{fig:annihilation_matrix}
\end{figure}

Suppose the device contains $L$ number of time bins (including the register) and the initial state is a $N$-photon Fock state. The Hilbert space $\mathcal{H}$ which spans the possible evolutions of this state can thought of as a system of $L$ entangleable $(N+1)$-qudits (since the vacuum state $\vac$ must additionally be supported). In this representation the Hilbert space has dimensionality $(N+1)^{L}$, and is spanned by the creation operators $\hat{a}_\ell^{\dagger^n}$, which create $0 \le n \le N$ photons in time bin $1 \le \ell \le L$. However, this representation is highly degenerate, supporting many invalid states, such as $\bigotimes_{\ell=1}^L \hat{a}_\ell^{\dagger (N-1)} \vac$, which has more than $N$ total photons in the system. To simulate even the modest $3 \times 3$ lattice depicted in Figure 1 of the main text requires operators with dimensionality $59049 \times 59049 \approx 3.5 \times 10^9$. 

A more tractable representation can be obtained by inverting the indexing and considering the possible placements of identical bosons into time bins. A number of identical bosons $N$ can be placed into $L$ distinct lattice sites in $\mathcal{N}^L_N = {{N + L - 1} \choose {L}}$ different ways. From this, we can generate a unique labeling of the allowable states in the system using a procedure devised by Ponomarev \cite{Ponomarev2011ThermalSystems, Ponomarev2009Ac-drivenMotor} and described in Ref. \cite{Raventos2017ColdDiagonalization}. We use the fact that $\mathcal{N}^L_N$ can be generated recursively as $\mathcal{N}^L = \sum_{n=0}^N \mathcal{N}^{L-1}_{N-n}$ to generate a counting scheme which evaluates all $\mathcal{N}^\ell_n$ up to $L,N$.

Suppose we have Fock state $\ket{\psi}$ with the occupancies of the $L$ lattice sites described in an $N$-dimensional vector $\left(\ell_1, \ell_2, \cdots , \ell_N \right)$, with $m_i \le m_j$ for $i>j$. The integer label $n_\psi$ of this state is 
\begin{equation}
n_\psi = 1 + \sum_{i=1}^N \mathcal{N}_i^{L-\ell_i}.
\end{equation}

For example, as shown in Figure \ref{fig:fock_to_ponomarev}, with $L=8, N=4$, the Fock state $\ket{\psi} = \ket{0,0,2,0,1,0,0,1}$ maps to $(\ell_1, \ell_2, \ell_3, \ell_4) = (8,5,3,3)$, and the integer label is $n_\psi = 1 + \mathcal{N}^3_1 + \mathcal{N}^3_2 + \mathcal{N}^5_3 + \mathcal{N}^8_4 = 112$. The inverse mapping is found with an iterative process: given $n_\psi$, we find the largest $\ell_N$ such that $\mathcal{N}^{\ell_N}_N < n_\psi$, then find the largest $\ell_{N-1}$ such that $\mathcal{N}^{\ell_{N-1}}_{N-1} < n_\psi - \mathcal{N}^{\ell_N}_N$, and so on.

To construct annihilation operators, we consider the direct sum of all smaller Hilbert spaces $\mathcal{H}^L_{n\le N} = \left[ \bigoplus_{n=0}^N \mathcal{H}^L_n \right]$. The annihilation operators $\hat{a}_i$ are constructed from a $\left(\sum_n \mathcal{N}^L_n \times \sum_n \mathcal{N}^L_n\right)$-dimensional matrix of zeros by iterating over all basis state labels $n_\psi = 1, 2, \cdots , \mathcal{N}^L_N$. To construct $\hat{a}_i$, for each Fock state $\ket{k_1 k_2 \cdots k_i \cdots k_L}$ with $k_i > 0$, a transition element is added which maps $\ket{k_1 k_2 \cdots k_i \cdots k_L} \mapsto \sqrt{k_i} \ket{k_1 k_2 \cdots (k_i - 1) \cdots k_L}$. The corresponding creation operator $\hat{a}^\dagger_i$ is simply the Hermitian conjugate of this operator, and photon expectation values in time bin $i$ are still $\langle \hat{a}^\dagger_i \hat{a}_i \rangle$ in this representation. The matrices are converted to \texttt{QuTiP} operators and the rest of the simulation is carried out normally. This process is illustrated in Figure \ref{fig:annihilation_matrix}.

This representation is especially effective for simulating systems with many more lattice sites than bosons $L \gg N$. For a $10 \times 10$ lattice containing two photons, the dimensionality of the state vectors is reduced from $5 \times 10^{47}$ to $5050$.

\section{Details for band structure computation}

\begin{figure}[t]
\centering 
\includegraphics[width=.65\textwidth]{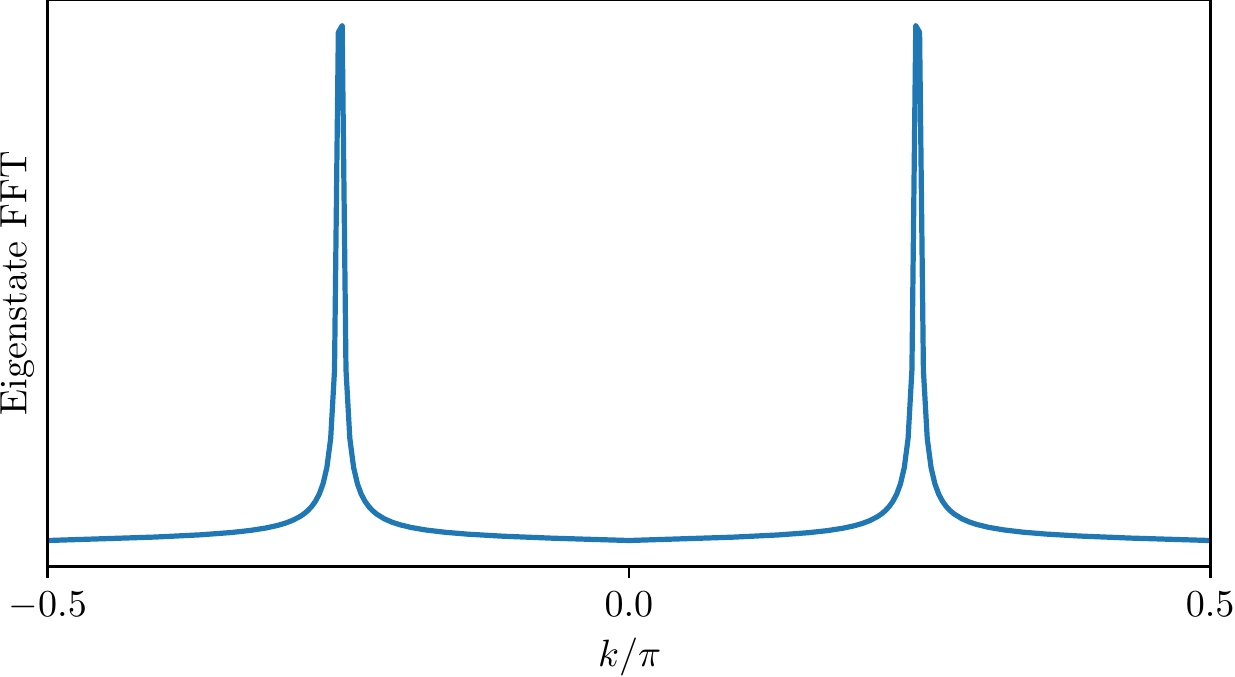}
\caption{
Computing the $k$ values for a lattice eigenstate using peak detection of the Fourier transform.
}
\label{fig:eigenstate_fft}
\end{figure}

The band structures in Fig. 2 of the main text are computed for the synthetic case by simulating one iteration of the propagator $\hat{G} = e^{-i \hat{H} (t=1)}$ in the device, taking the matrix logarithm $\hat{H} = \frac{\log \hat{G}}{-i}$, and then diagonalizing $\hat{H}$.  In both the emulated and exact cases, the Hamiltonian is represented in real space in the computer simulation; for each eigenstate of $\hat{H}$ with eigenvalue $E$, we compute $k$ values using peak detection of its Fourier transform, as shown in Figure \ref{fig:eigenstate_fft}. This can result in small numerical instabilities which are present in both the exact and emulated cases. Cases where the Fourier transform does not have clear peaks may result in outlying points with errant values of $k$; highly outlying points in Figure 2(b) of the main text have been pruned.

\section{Dispersion analysis}
\label{sec:emulator_dispersion_analysis}

In the linear regime of small $U$ (where we can ignore the effects of nonlinear pulse broadening), if we assume the same initial pulse shape with pulse length $\delta x$ within each bin of size $\Delta x$ (with $\delta x \ll \Delta x$), and that the register and storage loops are made of the same fiber, then all pulses will deform equally over time, so pulse distinguishability is not an issue, and a full-wave analysis such as the one presented in Ref. \cite{Bartlett2021DeterministicDimension} is not needed. The limiting factor will therefore be the dispersive length over which a pulse will broaden to the point where $\delta x \sim \Delta x$. Commercially available dispersion-shifted fibers allow for simultaneous low attenuation of $\sim 0.2$ dB/km and low dispersion of $\sim 4$ ps/nm/km at $\lambda = 1550$ nm. If we assume a bin size of $\Delta x = 20$ cm (time bin size of $1$ ns), a pulse length of $\delta x = 2$ cm, and we saturate the uncertainty limit $\Delta \lambda = \lambda^2 / 4 \pi \delta x$, then this dispersive length is approximately $26000$ km of fiber. At these distances, attenuation losses would certainly dominate, so we can safely ignore dispersive errors.


%